\documentclass[twocolumn, prl, superscriptaddress]{revtex4-1}
\usepackage{amssymb}
\usepackage{amsmath}
\usepackage{epsfig}
\usepackage{graphics, graphicx}
\usepackage{bbold}
\usepackage{psfrag}
\usepackage{mathcomp}
\usepackage{subfigure}
\usepackage{verbatim}
\usepackage{color}
\usepackage[colorlinks,citecolor=blue]{hyperref}
\def\cp#1{\mathbf{#1}}

\begin{document}

\title{Emergence of Crystalline Few-body Correlations in mass-imbalanced Fermi Polarons}

\author{Ruijin Liu}
\affiliation{Beijing National Laboratory for Condensed Matter Physics, Institute of Physics, Chinese Academy of Sciences, Beijing, 100190, China}
\author{Cheng Peng}
\affiliation{Beijing National Laboratory for Condensed Matter Physics, Institute of Physics, Chinese Academy of Sciences, Beijing, 100190, China}
\affiliation{School of Physical Sciences, University of Chinese Academy of Sciences, Beijing 100049, China}
\author{Xiaoling Cui}
\email{xlcui@iphy.ac.cn}
\affiliation{Beijing National Laboratory for Condensed Matter Physics, Institute of Physics, Chinese Academy of Sciences, Beijing, 100190, China}
\affiliation{Songshan Lake Materials Laboratory, Dongguan, Guangdong 523808, China}
%\date{August 24, 2020}

%\date{}

%\baselineskip24pt

\maketitle

\section{SUMMARY}
Polarons can serve as an ideal platform to identify few-body correlations in tackling complex many-body problems. In this work, we reveal various crystalline few-body correlations smoothly emergent from the mass-imbalanced Fermi polarons in two dimensions. A unified variational approach up to three particle-hole excitations allows us to extract the dominant dimer, trimer or tetramer correlation in a single framework. When the fermion-impurity mass ratio is beyond certain critical value, the Fermi polaron is found to undergo a smooth crossover, instead of a sharp transition, from the polaronic to trimer and tetramer regimes as increasing the fermion-impurity attraction. The emergent trimer and tetramer correlations result in the momentum-space crystallization of particle-hole excitations featuring a stable diagonal or triangular structure, as can be directly probed through the density-density correlation of majority fermions. Our results shed light on the intriguing quantum phases in the mass-imbalanced Fermi-Fermi mixtures beyond the pairing superfluid paradigm.

\begin{figure}[t]
\includegraphics[width=9cm]{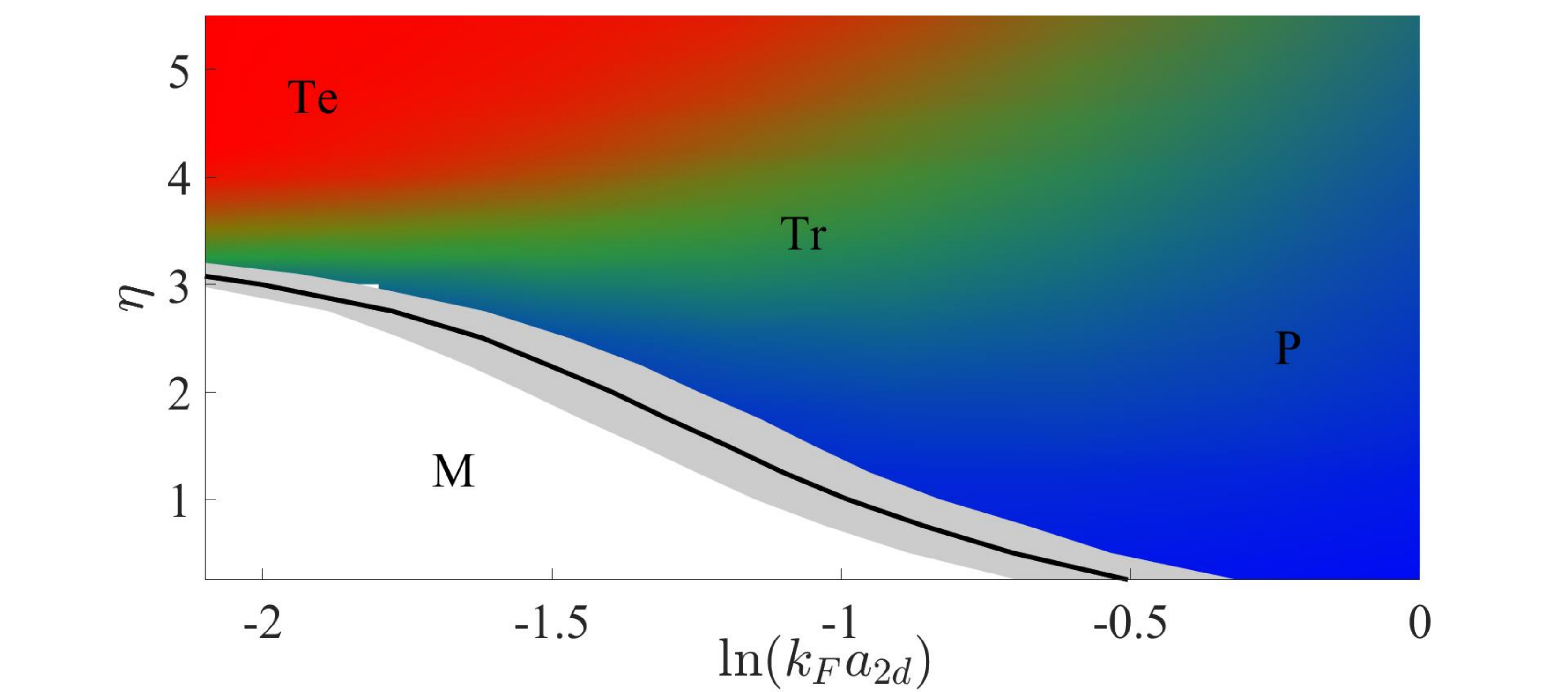}
\caption{(Color Online). Phase diagram of the 2D Fermi polaron with interaction strength $\ln(k_Fa_{2d})$ and mass ratio  $\eta\equiv m_f/m_{\rm a}$. The black line refers to the polaron-molecule phase boundary for small $\eta(<\eta_{tr})$: the ground state above the line is a polaron('P') with total momentum ${\cp Q}=0$ and below the line is a molecule('M') with $|{\cp Q}|=k_F$ (or with center-of-mass momentum $Q_M=0$).  The gray area around the line denotes the coexistence region for polaron and molecule. For $\eta>\eta_{tr}$, {\it no} phase transition is found and the ground state is always at ${\cp Q}=0$; as increasing the attraction strength, the system undergoes a smooth polaron-trimer('Tr') crossover for $\eta\in(\eta_{tr},\eta_{te})$ and a polaron-trimer-tetramer('Te') crossover for $\eta>\eta_{te}$.
For the ${\cp Q}=0$ ground state, we provide the RGB color map according to the weights of different particle-hole excitation terms in its full wave-function, i.e.,  $w_3$, $w_2$ and $w_1+w_0$ represent the mixing ratio of R (\textit{red}), G (\textit{green}) and B (\textit{blue}) colors, respectively.
Here  $\eta_{tr}=3.34$ and $\eta_{te}=3.38$ are, respectively, the critical mass ratios in 2D to stabilize a trimer and tetramer ground state in the few-body sector.  % The green, ..., and .. colors denote, respectively,    sequence of smooth crossover from polaronic (P) state to the dressed trimer (Tr) and tetramer (Te) states.
}  \label{fig_phase}
\end{figure}

\section{INTRODUCTION}

Many-body problems are usually complex, and an efficient tool to tackle them is through identifying few-body correlations.
A famous example is the Cooper instability, which tells the molecule formation on top of a spin-1/2 Fermi sea, providing the foundation for the theory of Cooper pair superfluidity\cite{book}. However, the hardest part of the approach is to foresee the dominant correlation from a complicated many-body environment. In this regard, the $1+N$ polaron system, which consists an impurity immersed in a sea of majority atoms, serves as an ideal platform for this study as it just interpolates between the few- and many-body systems and constitutes the simplest model for the few-to-many crossover\cite{Jochim_expt,polaron_review, polaron_review2}.

The concept of polaron, as raised by Landau nearly a century ago for an electron moving in solid\cite{Landau}, has been successfully applied to various physical systems. % such as high-Tc superconductors\cite{Mott}, semiconductors\cite{Lindemann}, neutral stars\cite{Kutschera}, etc.
In recent years, the polaron study has gained rapid development in ultracold atoms thanks to the high controllability over spin species, number, and interaction therein. For instance, by choosing majority atoms with different statistics, the cold atoms experiments have successfully realized the Fermi polaron\cite{Zwierlein,Salomon,Grimm,Kohl,Grimm2016,Roati,Zwierlein2,Sagi,Folling}, Bose polaron\cite{Arhus,JILA,Germany,MIT} and even both of them in a single system\cite{Grimm2021}, where the attractive and repulsive branches of polaron spectra have been explored (see recent reviews\cite{review_atom, review_rep}). Similar polaron spectra have also been observed in semiconductor microcavities\cite{exciton_polaron_expt1,exciton_polaron_expt2}.
Among all circumstances, the few-body correlation in an attractive Fermi polaron is of particular interest. It has been shown that the 3D and 2D Fermi polarons with equal mass can undergo a first-order transition from polaronic to molecular phase as  increasing the impurity-fermion attraction\cite{Prokofev1,Prokofev2,Punk,Leyronas,Bruun,Enss,Castin,Parish,MC_2d_1,MC_2d_2}, a consequence of  enhanced two-body correlation therein. Using a unified treatment of polaron and molecule, such transition can be interpreted as the energy competition between different momentum states\cite{Cui2020,Cui2021}, where the molecule serves as a good approximation for the finite-momentum state in strong coupling limit (see also \cite{ZhangYi}) and possesses a huge hidden degeneracy. The inclusion of finite-momentum states\cite{Cui2020,Cui2021,Parish_finiteT} is crucial for explaining the smooth polaron-molecule evolution in realistic experiment with a finite impurity density and at finite temperature\cite{Zwierlein,Kohl,Sagi}. Besides, the polaron-molecule transition has profound implications to the property of
 highly-polarized spin-1/2 fermions, where a
 phase separation between normal and superfluid states can occur\cite{PS_theory} as observed in cold atoms experiments\cite{PS_expt_Rice,PS_expt_MIT}.

The recent realization of mass-imbalanced Fermi-Fermi mixtures in cold atoms, such as  $^{40}$K-$^{6}$Li\cite{K_Li1, K_Li2, K_Li3}, $^{161}$Dy-$^{40}$K\cite{Dy_K1, Dy_K2}, $^{173}$Yb-$^{6}$Li\cite{Yb_Li, Yb_Li2} and $^{53}$Cr-$^{6}$Li\cite{Cr_Li, Cr_Li2}, offers an unprecedented
opportunity for uncovering intriguingly new correlation effect in Fermi polarons. Here an important hint is from the few-body physics, where the mass-imbalance has been found to greatly facilitate the formation of cluster bound states.
For instance, a light atom can bind with two heavy fermions to form an Efimov trimer\cite{Efimov} or a universal trimer\cite{KM,Pricoupenko}, and bind with three or four fermions to form a tetramer\cite{tetramer_efimov,universalTetramer,trimer_2d, pentemer, Peng} or pentamer\cite{pentemer,Peng}, as long as the heavy-light mass ratio is beyond certain critical value. An important follow-up question is how would their associated few-body correlations affect the many-body physics, such as, in $1+N$ Fermi polarons.  %, say, the Fermi polaron system.
In literature, there have been a number of  studies on the competition and conversion between polaron and trimer in Fermi polarons\cite{Parish_mass,Parish2,Nishida,Cui1,Cui2,Chevy2,Chevy3}.
However, a clear understanding on the relation between polaron, trimer and other highly correlated states is still missing, due to the lack of a unified perspective on them. It is still open questions how the different few-body correlations evolve and manifest themselves in polaron systems.
The problem is quite challenging as it requires a theoretical approach to incorporate all essential correlations in a single framework.

In this work, we address the mass-imbalanced Fermi polaron problem in 2D using a unified variational approach consisting up to three particle-hole excitations. The approach represents the state-of-the-art theoretical tool that allows us to systematically examine the $n$-body correlations, with $n$ ranging from $2$ (dimer), $3$ (trimer) to $4$ (tetramer), in a single unified framework. We find that in distinct contrast to the first-order transition in equal mass case, {\it no} sharp transition occurs in the mass-imbalanced Fermi polaron if the fermion-impurity mass ratio $\eta$ is larger than $\eta_{tr}(=3.34)$, the critical value to support a trimer bound state in the few-body sector.
Depending on the actual $\eta$($>\eta_{tr}$), the system undergoes a continuous evolution from a polaronic state to the dressed trimer or tetramer states (see Fig.\ref{fig_phase}), where the dominant $3$- and $4$-body correlations gradually emerges. In this process, the majority fermions develop the {\it momentum-space crystallization}, i.e., the dominant particle-hole excitations distribute with equal interval near the Fermi surface that form a stable diagonal or triangular structure, despite the whole system is rotational invariant. Such emergent crystallization directly characterizes the dominant trimer and tetramer correlations, and can be readily detected via the density-density correlation of majority fermions.
Our results reveal a remarkable effect of  mass imbalance in changing the few-body correlation in Fermi polarons, which  shed light on the novel phases and correlations in mass-imbalanced Fermi-Fermi mixtures that are well beyond the pairing superfluid paradigm.

 In the rest part of the paper, we will present the unified variational approach for Fermi polaron problem, as well as the results of polaron-trimer or polaron-trimer-tetramer crossover for the mass-imbalanced case with emergent crystalline few-body correlations. The final discussion part will be devoted to the summary, generalization and  implication of our results in a broader context.

\section{RESULTS}

\subsection{Unified variational approach}

We start from the Hamiltonian describing a single impurity immersed in a 2D Fermi sea:
\begin{equation}
H=\sum_{\mathbf{k}} \left(\epsilon^a_{\mathbf{k}} a_{\mathbf{k}}^{\dagger} a_{\mathbf{k}} + \epsilon^f_{\mathbf{k}} f_{\mathbf{k}}^{\dagger} f_{\mathbf{k}}\right) +\frac{g}{S} \sum_{\mathbf{Q}, \mathbf{k}, \mathbf{k}^{\prime}} a_{\mathbf{Q}-\mathbf{k}}^{\dagger} f_{\mathbf{k}}^{\dagger} f_{\mathbf{k}^{\prime}} a_{\mathbf{Q}-\mathbf{k}^{\prime}}. \label{eq:H}
\end{equation}
Here $a_{\mathbf{k}}^{\dagger}$ and $f_{\mathbf{k}}^{\dagger}$ respectively create an impurity atom and a majority fermion with momentum ${ \mathbf{k}}$ and energy $\epsilon^{a,f}_{\mathbf{k}}=\mathbf{k}^2/(2m_{a,f})$; $g$ is the bare coupling that follows the renormalization equation $1/g=-1/S \sum_{\mathbf{k}}1/(\epsilon^a_{\mathbf{k}}+\epsilon^f_{\mathbf{k}}+E_{2b})$, where $S$ is the system area and $E_{2b}=(2\mu a_{2d}^2)^{-1}$ is the 2-body binding energy in vacuum that relies on the relative mass $\mu=m_am_f/(m_a+m_f)$ and 2D scattering length $a_{2d}$. We take $\hbar=1$ for brevity. The Fermi polaron properties are determined by two dimensionless parameters, namely, the mass ratio $\eta\equiv m_f/m_a$ and the interaction strength $\ln(k_Fa_{2d})$, where $k_F$ is the Fermi momentum of majority atoms giving the Fermi energy $E_F=k_F^2/(2m_f)$.

We write down a generalized version of the Fermi polaron ansatz  initially proposed in \cite{Chevy}:
%\begin{eqnarray}
%P_{2n+1}({\mathbf{Q}})=\left(\psi^{(0)} a^{\dag}_{{ \mathbf{Q}}}+\frac{1}{(n!)^2}\sum_{l=1}^n \sum_{{ \mathbf{k}}_i { \mathbf{q}}_j}\psi^{(l)}_{{ \mathbf{k}}_i { \mathbf{q}}_j} a^{\dag}_{{ \mathbf{P}}}\prod_{i=1}^l f^{\dag}_{{ \mathbf{k}}_i}\prod_{j=1}^l f_{{ \mathbf{q}}_j}\right) |{\textrm {FS}}\rangle_{ N}. \label{p_Q}
%\end{eqnarray}
\begin{align}
&P_{2n+1}({\cp Q})=\left(\psi^{(0)} a^{\dag}_{{\cp Q}}+\right. \nonumber\\
&\ \ \ \ \ \ \ \left.\frac{1}{(n!)^2}\sum_{l=1}^n \sum_{{\cp k}_i {\cp q}_j}\psi^{(l)}_{{\cp k}_i {\cp q}_j} a^{\dag}_{{\cp P}}\prod_{i=1}^l f^{\dag}_{{\cp k}_i}\prod_{j=1}^l f_{{\cp q}_j}\right) |{\rm FS}\rangle_{N}. \label{p_Q}
\end{align}
Here $|\textrm {FS}\rangle_{N}$ is the Fermi sea with number $N$; all ${ \mathbf{q}}$ (${ \mathbf{k}}$) are below (above) the Fermi surface and ${ \mathbf{P}}={\mathbf{Q}}+\sum_j{\mathbf{q}}_j-\sum_i{\mathbf{k}}_i$. Taking $|\textrm {FS}\rangle_{N}$ as the reference state, $P_{2n+1}({\mathbf{Q}})$ describes the impurity dressed with up to $n$ particle-hole(p-h) excitations and with total momentum ${\mathbf{Q}}$. Because of the rotational invariance, hereafter we will use $Q\equiv|{\mathbf{Q}}|$ as the momentum index for simplicity.

We emphasize that the generalized ansatz Eq. \ref{p_Q} systematically incorporates all the essential few-body correlations and thus can serve as a unified approach for the Fermi polaron problem. First, taking $Q=0$, it describes the polaronic ground state  for the weak coupling Fermi polarons, as studied previously with up to one\cite{Chevy,Combescot1,Punk,Castin,Parish,Parish_mass,Pethick,Cui2020} and two\cite{Combescot2,Leyronas,Parish2,Cui2021} particle-hole (p-h) excitations. Secondly, taking a finite $Q=k_F$, it well covers the molecule state with zero center-of-mass momentum as proposed in the strong coupling regime\cite{Punk,Leyronas,Parish,Castin,Parish_mass,Parish2,Pethick}, where the impurity essentially binds with one fermion on the Fermi surface to form a dimer. In this sense, it has been pointed out that the polaron-molecule transition can be recognized as the energy competition between different $Q$-sectors\cite{Cui2020,Cui2021}. Finally, by including higher-order p-h excitations, all the few-body correlations can be systematically incorporated. Specifically, the full $n$-body correlation effect has been contained in the $(n-1)$ p-h excitation terms in Eq. \ref{p_Q}.

In this work, we will consider up to three p-h  excitations, i.e., with $P_7({Q})$, which allow us to unify the dimer, trimer, and tetramer correlations in a single framework to examine their relations and competitions in-between. See note S1 for the integral equations from $P_7({Q})$ and the numerical details are presented in note S2, Figs.S1 and S2. Moreover, since the $P_7$ formulism can well cover the $P_5$ ones by neglecting the terms of three p-h excitations, in the regime where these excitations (signifying tetramer correlation) are negligible we shall work with $P_5$ for simplification.
Finally, we remark that there are two important values of mass ratio to distinguish different few-body correlations  in 2D Fermi polarons studied here, namely, the one to support a trimer in vacuum from the atom-dimer threshold $\eta_{tr}=3.34$\cite{Pricoupenko} and the one to support a tetramer in vacuum from the atom-trimer threshold  $\eta_{te}=3.38$\cite{Peng}. The effect of even larger cluster bound states will be discussed at the end of this paper.

\subsection{Polaron-molecule transition for $\eta<\eta_{tr}$}

\begin{figure}[h]
\includegraphics[width=8.8cm]{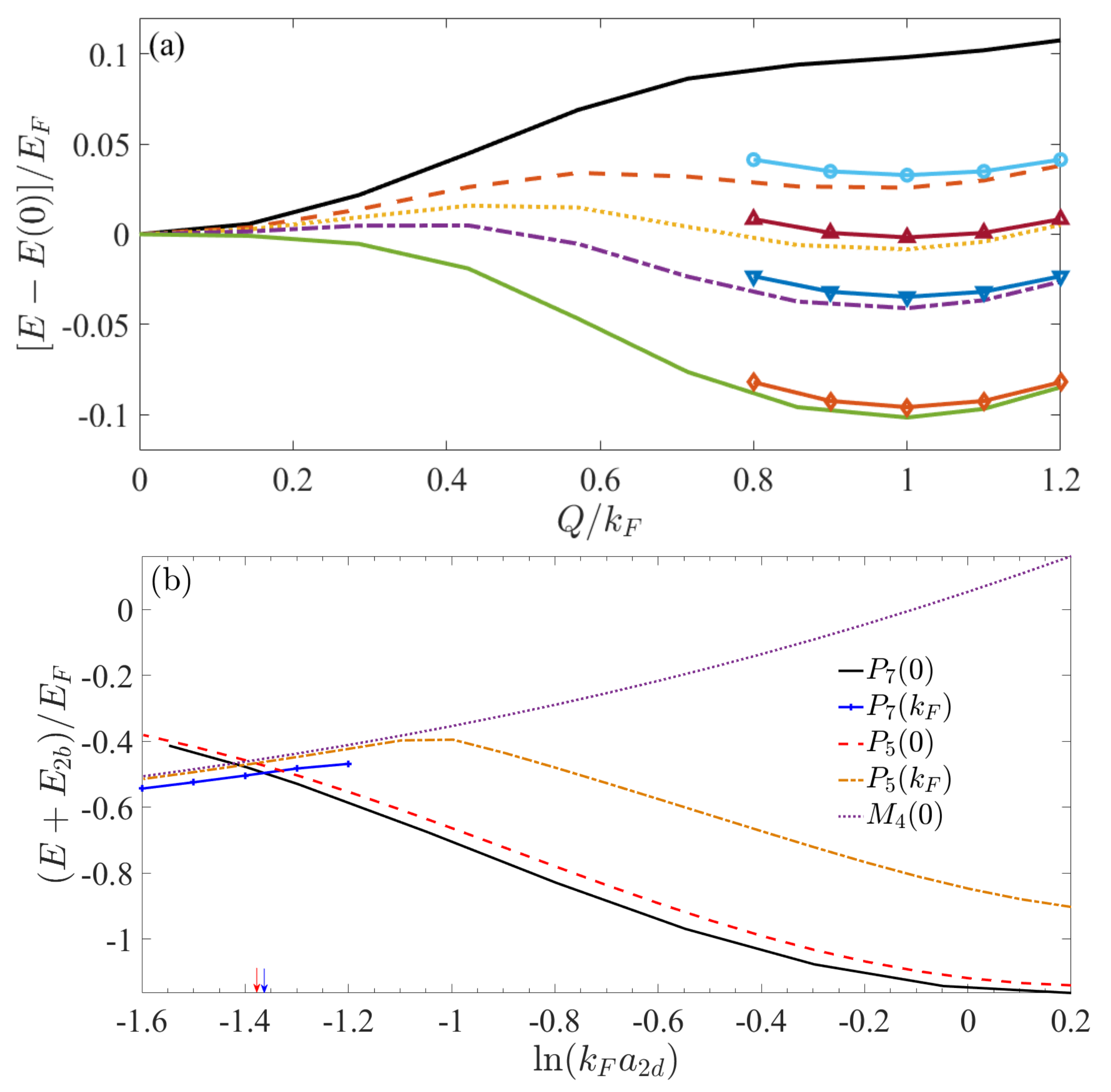}
\caption{(Color Online). Polaron-molecule transition at mass ratio $\eta=2$. (a) Energy dispersion from $P_5(Q)$ for interaction strength $\ln(k_Fa_{2d})=-1.25, -1.35, -1.4, -1.45, -1.55$ (from top to bottom), shifted by the energy at $Q=0$. The lines with points show the energies from $M_4({Q}_M)$ after a constant momentum shift $k_F$. (b) Energy comparison between $P_7(0)$, $P_7(k_F)$, $P_5(0)$, $P_5(k_F)$ and $M_4(0)$. For all coupling regime, $P_7(Q)$ and $P_5(Q)$ (with $Q=0,\ k_F$) produce very close energies (with $\sim 0.03 E_F$ deviation at most), indicating the negligible role of three p-h excitations in this case.  In strong coupling regime, $M_4(0)$ can well approximate $P_5(k_F)$ and their energies are indistinguishable. The polaron-molecule transition locates at $\ln(k_Fa_{2d})=-1.36$(blue arrow) as given by the energy crossing between $P_7(0)$ and $P_7(k_F)$, very close to the transition point $-1.38$ (red arrow) as given by the crossing between $P_5(0)$ and $P_5(k_F)$.  }  \label{fig_PMtransition}
\end{figure}

We first briefly go through the regime with a small mass ratio $\eta<\eta_{tr}$. In this case, the three p-h excitations in $P_7$ take little effect and thus $P_5$ can serve as a good ansatz. In this regime the polaron-molecule transition persists, with the same nature as in equal mass case. Taking $\eta=2$ for instance, we show in Fig.\ref{fig_PMtransition}A that as increasing the coupling strength, the ground state of Fermi polaron can switch from $Q=0$ to $Q=k_F$, signifying the polaron to molecule transition. Nearby the transition, the dispersion curve displays a double well structure where $Q=0$ and $Q=k_F$ states are both locally stable, indicating the polaron-molecule coexistence in realistic system with a finite impurity density and at finite temperature. All these features are the same as equal mass case\cite{Cui2020, Cui2021}.

For comparison, we also consider the molecule ansatz
%\begin{eqnarray}
%M_{2n+2}({ \mathbf{Q}}_{M})=\left(\sum_{{ \mathbf{k}}}\phi^{(0)}_{{ \mathbf{k}}} a^{\dag}_{{ \mathbf{Q}_{M}}-{ \mathbf{k}}}f^{\dag}_{ \mathbf{k}}+\sum_{l=1}^n \sum_{\mathbf{k}{ \mathbf{k}}_i { \mathbf{q}}_j}\phi^{(l)}_{{ \mathbf{k}}{ \mathbf{k}}_i { \mathbf{q}}_j} a^{\dag}_{{ \mathbf{P}}}f^{\dag}_{{ \mathbf{k}}} \prod_{i=1}^l f^{\dag}_{{ \mathbf{k}}_i}\prod_{j=1}^l f_{{ \mathbf{q}}_j}\right) |{\textrm {FS}}\rangle_{N-1},\label{wf_m}
%\end{eqnarray}
\begin{align}
&M_{2n+2}({\cp Q}_{M})=\left(\sum_{{\cp k}}\phi^{(0)}_{{\cp k}} a^{\dag}_{{\cp Q_{M}}-{\cp k}}f^{\dag}_{\cp k}+\right. \nonumber\\
&\ \ \ \ \ \ \left.\sum_{l=1}^n \sum_{\mathbf{k}{\cp k}_i {\cp q}_j}\phi^{(l)}_{{\cp k}{\cp k}_i {\cp q}_j} a^{\dag}_{{\cp P}}f^{\dag}_{{\cp k}} \prod_{i=1}^l f^{\dag}_{{\cp k}_i}\prod_{j=1}^l f_{{\cp q}_j}\right) |{\rm FS}\rangle_{N-1},\label{wf_m}
\end{align}
with ${ \mathbf{P}}={\mathbf{Q}}_{M}-{\mathbf{k}}-\sum_{i=1}^n ({ \mathbf{k}}_i-{ \mathbf{q}}_i)$. As pointed out in Ref.\cite{Cui2020,Cui2021}, $M_{2n+2}(0)$ expands a subset of the variational space of $P_{2n+3}(k_F)$, and thus the former is always energetically unfavorable as compared to the latter. However, in the strong coupling regime, $M_{2n+2}(0)$ can be a good approximation for $P_{2n+3}(k_F)$. As shown in Fig.\ref{fig_PMtransition}B for the case of $\eta=2$, at the transition between $Q=0$ and $Q=k_F$, the energies from $M_4(0)$ and $P_5(k_F)$ are indistinguishable and thus it can be interpreted as the polaron-molecule transition. In this case, since $P_5(Q)$ produce very close energies with $P_7(Q)$, these two ansatz result in very similar critical points for polaron-molecule transition, see the blue and red arrows in Fig.\ref{fig_PMtransition}B.

In addition, we would like to clarify that in strong coupling regime, $P_7(0)$ ansatz can produce another type of solution that has very close energy with $M_4(0)$. This solution corresponds to considering a special structure of p-h excitations near the Fermi surface, and thus can be approached in our numerics by choosing a proper initial state in the iteration. However, such state is only stabilized when the interaction is strong enough, and it is always orthogonal to the one that is adiabatically evolved from the weak coupling limit (as shown by the black line in Fig.\ref{fig_PMtransition}B). Moreover, this state belongs to an excited manifold, i.e., with a higher energy than $P_7(k_F)$, in the strong coupling regime, due to the fact that $P_7(k_F)$ is well approximated by $M_6(0)$ while this solution is close to $M_4(0)$. Therefore, with a fixed truncated level of p-h excitations, the lowest molecule state is given by $P_{2n+1}(k_F)$, rather by a special solution of $P_{2n+1}(0)$. In this sense, the presence of this solution will not affect the occurrence and the nature of polaron-molecule transition in this system. More details on this solution can be found in note S3 and Fig.S3.

In Fig.\ref{fig_phase}, we have mapped out the transition point and the coexistence region for polaron ($Q=0$) and molecule ($Q=k_F$, $Q_M=0$)  in ($\eta,\ \ln(k_Fa_{2d})$) plane. It is found that as increasing $\eta$, the critical $\ln(k_Fa_{2d})_c$ moves to strong coupling regime and saturates at $-\infty$ as $\eta\rightarrow\eta_{tr}$. This is consistent with the ground state in the vacuum limit ($k_F\rightarrow 0$), where the molecule (or dimer) gives way to a trimer ground state  at  $\eta_{tr}$ and the trimer gives way to tetramer at $\eta_{te}$. The phase diagram in Fig.\ref{fig_phase} exactly reflects such ground state change in $k_F\rightarrow 0$ limit ($\ln(k_Fa_{2d})\rightarrow -\infty$), where the trimer and tetramer regions are marked respectively by green and red colors. Increasing $k_F$ from zero, these cluster bound states will be significantly affected by the majority Fermi sea and finally, as discussed below, all evolve into the polaronic state at large $k_F$ ($\ln(k_Fa_{2d})\rightarrow \infty$).

\subsection{Smooth polaron-trimer crossover  for $\eta_{tr}<\eta<\eta_{te}$}

\begin{figure*}[t]
\includegraphics[width=18cm]{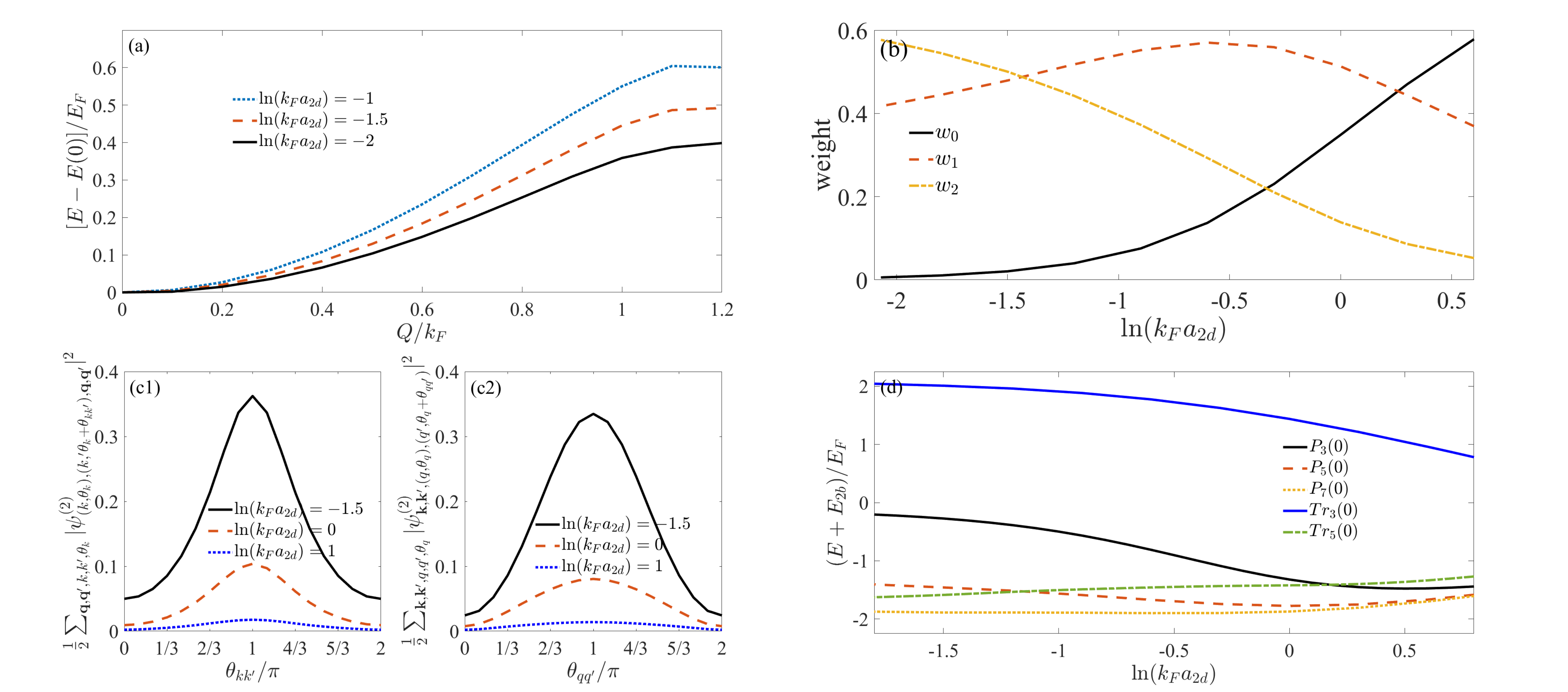}
\caption{(Color Online). Smooth polaron-trimer crossover  at mass ratio $\eta=3.35$. (a) Energy dispersion from $P_5(Q)$ for different interaction strengths $\ln(k_Fa_{2d})=-1, -1.5, -2$(from top to bottom), shifted by the energy at $Q=0$. (b) Weights of the bare ($w_0$), one p-h ($w_1$), and two p-h ($w_2$) terms in $P_5(0)$ as functions of $\ln(k_Fa_{2d})$. (c1 or c2) Probability of finding two particles (or two holes) with relative angle $\theta_{{\cp k}{\cp k}'}$ (or $\theta_{{\cp q}{\cp q}'}$) for different couplings $\ln(k_Fa_{2d})=-1.5, 0, 1$(from top to bottom).  %and two holes   Upper panel: $\sum_{{\cp k}{\cp k}'}|\psi^{(2)}_{\mathbf{kk'qq'}}|^2$ as a function of the relative angle of two holes $\theta_{{\cp q}{\cp q}'}$ with $|{\cp q}|=|{\cp q}'|=0.9...k_F$, signifying the hole-hole angular correlation. Lower panel: $\sum_{{\cp q}{\cp q}'}|\psi^{(2)}_{\mathbf{kk'qq'}}|^2$ as a function of relative angle of two particles $\theta_{{\cp k}{\cp k}'}$, with $|{\cp k}|=|{\cp k}'|=1.0...k_F$, signifying the particle-particle angular correlation.
%The inset shows the density-density correlation of majority fermions in momentum space $D(\mathbf{k})\equiv \langle n^f_{\cp k_0}n^f_{\cp k}\rangle$, with a fixed $\mathbf{k}_0=()k_F$.
(d)Energy comparison between $P_3(0)$, $P_5(0)$, $P_7(0)$, $Tr_3(0)$ and $Tr_5(0)$. %(e) Effective mass of $Q=0$ ground state, obtained from $P_7$ and $P_5$, as a function of $\ln(k_Fa_{2d})$.
} \label{fig_PTr_crossover}
\end{figure*}

A remarkable difference between Fermi polarons with $\eta<\eta_{tr}$ and $\eta>\eta_{tr}$ is that the latter displays {\it no} sharp transition as the impurity-fermion attraction increases. In order for a direct comparison with Fig.\ref{fig_PMtransition}A, in Fig.\ref{fig_PTr_crossover}A we show the same energy dispersion from $P_5(Q)$ but at a higher mass ratio $\eta=3.35(>\eta_{tr})$. In this case the ground state always stays at $Q=0$, without any transition to other $Q$-sectors.

To examine the inner structure of $Q=0$ state, we plot in Fig.\ref{fig_PTr_crossover}B the respective weight of the bare term ($w_0$),  one ($w_1$) and two p-h excitation terms ($w_2$), which are defined as
\begin{equation}
w_n= \frac{1}{(n!)^2}\sum_{\{{\mathbf{k}}\} \{{ \mathbf{q}}\}}|\psi^{(n)}_{{ \mathbf{k}}_1...{ \mathbf{k}}_n { \mathbf{q}}_1...{ \mathbf{q}}_n}|^2,  \label{wn}
\end{equation}
with the constraint $\sum_n w_n=1$ from the normalization. By definition, $w_0$ is exactly the residue of the polaron. We can see that as tuning the coupling $\ln(k_Fa_{2d})$ from weak to strong, the system evolves from a polaronic state where $w_0$ dominates, to $w_1$-dominated intermediate state and finally ends up  at the $w_2$-dominated state. For this final state, the polaronic picture fails given the residue $w_0\sim 0$, and its another important feature is that the two p-h excitations are strongly correlated near the Fermi surface. As shown in Figs.\ref{fig_PTr_crossover}C-D, the probability of finding two particles (and two holes) reaches a pronounced peak if their relative angle is $\pi$, i.e., they tend to distribute with opposite directions as a diagonal. Such diagonal distribution is a direct manifestation of trimer correlation, which describes an impurity near zero momentum bound with two particles with opposite momenta  outside the Fermi sea (associated with hole creations inside) to form a trimer. Therefore the smooth evolution in Fig.\ref{fig_PTr_crossover}B can be seen as the polaron-trimer crossover. In this process, the $w_1$-dominated intermediate state serves as a bridge connecting the polaron and trimer regimes.

Previously, a sharp transition between polaron and trimer has been concluded for the same mass-imbalanced system\cite{Parish_mass,Parish2}, in contrast to our finding here.
The discrepancy can be resolved by examining the relation between polaron and trimer ansatz, with the latter given by
%\begin{eqnarray}
%&Tr_{2n+3}({0})=\sum\limits_{{ \mathbf{k}}{ \mathbf{k}}'}\left(\tau^{(0)}_{{ \mathbf{k}}{ \mathbf{k}}'}a^{\dag}_{-\mathbf{k}-\mathbf{k}'}f^{\dag}_{{\mathbf{k}}}f^{\dag}_{{ \mathbf{k}}'} +\sum\limits_{l=1}^n \sum\limits_{{\mathbf{k}}_i {\mathbf{q}}_j}\tau^{(l)}_{{ \mathbf{k}}{ \mathbf{k}}'{ \mathbf{k}}_i { \mathbf{q}}_j} a^{\dag}_{{ \mathbf{P}}}f^{\dag}_{{ \mathbf{k}}}f^{\dag}_{{\mathbf{k}}'}\prod\limits_{i=1}^l f^{\dag}_{{ \mathbf{k}}_i}\prod\limits_{j=1}^l f_{{ \mathbf{q}}_j}\right) |{\rm FS}\rangle_{N-2}. \label{tr}
%\end{eqnarray}
\begin{align}
&Tr_{2n+3}({\cp 0})=\sum_{{\cp k}{\cp k}'}\left(\tau^{(0)}_{{\cp k}{\cp k}'}a^{\dag}_{-\mathbf{k}-\mathbf{k}'}f^{\dag}_{{\cp k}}f^{\dag}_{{\cp k}'} +\right. \nonumber\\
&\ \ \ \ \ \ \left.\sum_{l=1}^n \sum_{{\cp k}_i {\cp q}_j}\tau^{(l)}_{{\cp k}{\cp k}'{\cp k}_i {\cp q}_j} a^{\dag}_{{\cp P}}f^{\dag}_{{\cp k}}f^{\dag}_{{\cp k}'}\prod_{i=1}^l f^{\dag}_{{\cp k}_i}\prod_{j=1}^l f_{{\cp q}_j}\right) |{\rm FS}\rangle_{N-2}. \label{tr}
\end{align}
with ${\mathbf{P}}=-{\mathbf{k}}-{ \mathbf{k}}'-\sum_{i=1}^n ({ \mathbf{k}}_i-{ \mathbf{q}}_i)$. Clearly,  Eq. \ref{tr} describes a trimer outside the Fermi sea dressed with p-h excitations. The polaron-trimer transition was drawn from an energy crossing between $P_3(0)$ and $Tr_5(0)$\cite{Parish_mass,Parish2}.

Here, we remark that the polaron and trimer states are intimately related to each other. As indicated by Figs.\ref{fig_PTr_crossover}C-D,
the trimer ansatz $Tr_{2n+3}(0)$ (Eq. \ref{tr}) just corresponds to picking up  a particular set of p-h excitations in the polaron ansatz $P_{2n+5}(0)$ (Eq. \ref{p_Q}) where the two holes are right at the Fermi surface and with opposite directions. Therefore, $Tr_{2n+3}$ falls into the same $Q$-sector as $P_{2n+5}$, and actually belongs to a subset of the latter. Because of this, $Tr_{2n+3}(0)$ should be always energetically unfavorable as compared to $P_{2n+5}(0)$, and the energy of the former can only approach the latter from above but there cannot be an energy crossing. This is to say, the polaron to trimer evolution, if exists, can only be a crossover but not a sharp transition. Note that it is very different from the case of polaron-molecule competition, where $P_{2n+3}(0)$ and $M_{2n+2}(0)$ belong to different $Q$-sectors (taking $|\textrm{FS}\rangle_{N}$ as reference) and there can be a transition between them\cite{Cui2020,Cui2021}.
Moreover, we would like to clarify that the energy comparison between $Tr_{2n+3}$ and $P_{2n+1}$ (with the same $n$)  is problematic in identifying a transition, since the former contains two more particles outside the Fermi sea and their additional scattering can easily produce a lower energy than the latter. A proper comparison is between $Tr_{2n+3}$ and $P_{2n+5}$, both  having $n+3$ particles outside the Fermi sea.

In Fig.\ref{fig_PTr_crossover}E, we compare the energies from various ansatz for $\eta=3.35(>\eta_{tr})$. As expected, the energy of $Tr_5(0)$ (or $Tr_3(0)$) is visibly higher than that of $P_7(0)$ (or $P_5(0)$). Nevertheless, $Tr_5(0)$ does have an energy crossing with $P_3(0)$ and $P_5(0)$, essentially because they are actually associated with different levels of p-h excitations on top of the Fermi sea $|\textrm{FS}\rangle_{N}$. Such crossing cannot represent any transition in the system. Based on these analyses, we expect a similar polaron-trimer crossover instead of a transition in other Fermi polaron systems\cite{Cui2,Chevy3}.

\subsection{Polaron-trimer-tetramer crossover and crystalline correlation for $\eta>\eta_{te}$}

Continuously increasing mass ratio $\eta$ beyond $\eta_{te}$, the four-body (tetramer) correlation can become dominant and lead to even intriguing physics. To capture the physics here, we have worked with $P_7$ ansatz which includes up to three p-h excitations and thus covers all the dimer, trimer and tetramer correlations. We have confirmed the absence of phase transition in this case and the ground state is always at $Q=0$, similar to the case of $\eta\in(\eta_{tr},\eta_{te})$. However, here a crucial difference is that as increasing the coupling strength, the system undergoes a sequence of crossover from  polaron to trimer, and then from trimer to tetramer states.

To understand the relation between the tetramer and the general ansatz in Eq. \ref{p_Q}, we write down the tetramer ansatz according to the same strategy as in writing Eq. \ref{wf_m} and Eq. \ref{tr}:
%\begin{eqnarray}
%&Te_{2n+4}({ 0})=\sum\limits_{{ \mathbf{k}}{ \mathbf{k}}'{ \mathbf{k}}''}\left(\nu^{(0)}_{{ \mathbf{k}}{ \mathbf{k}}'{ \mathbf{k}}''}a^{\dag}_{-\mathbf{k}-\mathbf{k}'}f^{\dag}_{{ \mathbf{k}}}f^{\dag}_{{ \mathbf{k}}'}f^{\dag}_{{ \mathbf{k}}''} +\sum\limits_{l=1}^n \sum\limits_{{ \mathbf{k}}_i { \mathbf{q}}_j}\nu^{(l)}_{{ \mathbf{k}}{ \mathbf{k}}'{ \mathbf{k}}''{ \mathbf{k}}_i { \mathbf{q}}_j} a^{\dag}_{{ \mathbf{P}}}f^{\dag}_{{ \mathbf{k}}}f^{\dag}_{{ \mathbf{k}}'}f^{\dag}_{{ \mathbf{k}}''}\prod\limits_{i=1}^l f^{\dag}_{{ \mathbf{k}}_i}\prod\limits_{j=1}^l f_{{ \mathbf{q}}_j}\right) |{\textrm {FS}}\rangle_{N-3}, \label{te}
%\end{eqnarray}
\begin{align}
&Te_{2n+4}({\cp 0})=\sum_{{\cp k}{\cp k}'{\cp k}''}\left(\nu^{(0)}_{{\cp k}{\cp k}'{\cp k}''}a^{\dag}_{-\mathbf{k}-\mathbf{k}'}f^{\dag}_{{\cp k}}f^{\dag}_{{\cp k}'}f^{\dag}_{{\cp k}''} +\right. \nonumber\\
&\ \ \ \ \ \left.\sum_{l=1}^n \sum_{{\cp k}_i {\cp q}_j}\nu^{(l)}_{{\cp k}{\cp k}'{\cp k}''{\cp k}_i {\cp q}_j} a^{\dag}_{{\cp P}}f^{\dag}_{{\cp k}}f^{\dag}_{{\cp k}'}f^{\dag}_{{\cp k}''}\prod_{i=1}^l f^{\dag}_{{\cp k}_i}\prod_{j=1}^l f_{{\cp q}_j}\right) |{\rm FS}\rangle_{N-3}, \label{te}
\end{align}
with ${\mathbf{P}}=-{\mathbf{k}}-{ \mathbf{k}}'-{ \mathbf{k}}''-\sum_{i=1}^n ({ \mathbf{k}}_i-{ \mathbf{q}}_i)$. Eq. \ref{te} describes a tetramer bound state that is on top of the Fermi sea and dressed with p-h excitations. Then, following a similar analysis as in the trimer case, we can immediately see that $Te_{2n+4}$ belongs to a special case of $P_{2n+7}$ by only considering the p-h excitations with three holes pinning at the Fermi surface. By manipulating the directions of three holes such that their total momentum is zero, these two ansatz can belong to the same $Q$-sector and therefore $Te_{2n+4}(0)$ expands a sub-variational space of $P_{2n+7}(0)$. Because of these, the conversion between polaron and tetramer regimes can only be a smooth crossover but not a sharp transition, similar to the polaron-trimer crossover discussed earlier.

\begin{figure*}[t]
\includegraphics[width=18cm]{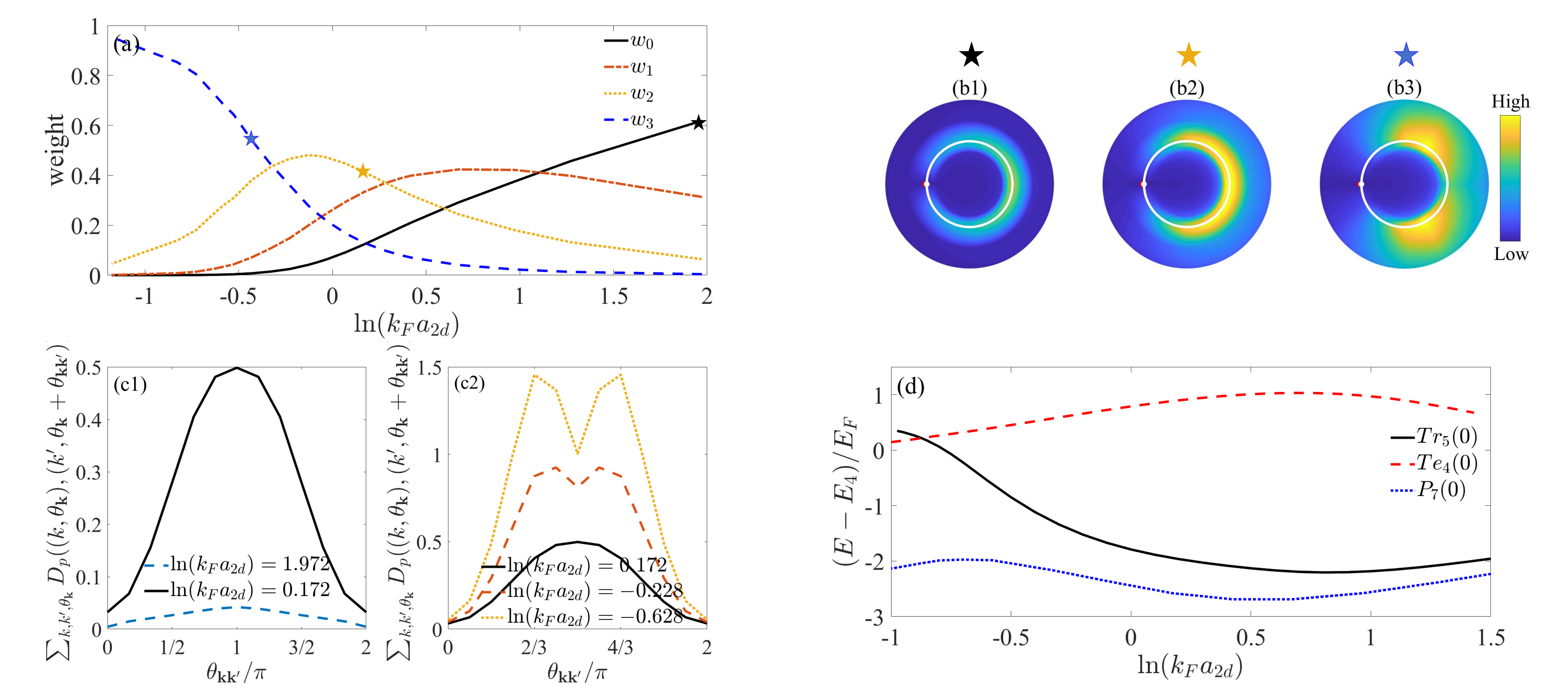}
\caption{(Color Online). Polaron-trimer-tetramer  crossover at mass ratio $\eta=40/6$. (a) Weights  of the bare term and various p-h excitation terms (see the $w_n$ definition in (\ref{wn})) in $P_7(0)$ as functions of $\ln(k_Fa_{2d})$.
% Energy comparison of $P_7(0)$, $Tr_5(0)$, and $Te_4(0)$.
(b1-b3) Contour plots of hole-hole correlation $D_h({\cp q_0},{\cp q})$ and particle-particle correlation $D_p({\cp k}_0,{\cp k})$ for different $\ln(k_Fa_{2d})=1.972(b1), 0.172(b2), -0.428(b3)$, which, respectively, belong to the polaron, trimer and tetramer regime as labeled by different stars in (a). Here we take ${\cp q}_0=-k_F{\cp e}_x$ and ${\cp k}_0=-1.08k_F{\cp e}_x$, as marked by white and red points in the plots. The white circle denotes the Fermi surface. (c1,c2) Particle-particle angular correlation during the crossover from polaron to trimer (c1), and from trimer to tetramer (c2). %Here we take $|{\cp k}|=|{\cp k}_0|=..$, and $\theta_{{\cp k}{\cp k}_0}$ is the angle between ${\cp k}$ and ${\cp k}_0$.
(d) Energy comparison between $P_7(0)$, $Tr_5(0)$ and $Te_4(0)$. %{\color{red} Here $E_4=-1.282E_{2b}$ is the energy of the bare tetramer.}
} \label{fig_PTrTe_crossover}
\end{figure*}

In Fig.\ref{fig_PTrTe_crossover}, we take the $^{40}$K-$^{6}$Li Fermi polaron with $\eta=40/6$ and demonstrated the polaron-trimer-tetramer crossover therein. Fig.\ref{fig_PTrTe_crossover}A shows that as increasing the coupling strength, the system undergoes a continuous evolution from a polaronic state (where $w_0$ dominates), to trimer ($w_2$ dominates) and finally to tetramer state ($w_3$ dominates). These states can also be distinguished from the momentum-space correlation of majority fermions.  To see this, we compute the hole-hole and particle-particle correlation functions of majority fermions in momentum space,  defined via
\begin{eqnarray}
D_h({ \mathbf{q}_0},{ \mathbf{q}})&\equiv& \langle (1-n^f_{\mathbf{q}_0})(1-n^f_{\mathbf{q}}) \rangle, \label{Dhh}\\
D_p({ \mathbf{k}_0},{ \mathbf{k}})&\equiv& \langle n^f_{ \mathbf{k}_0}n^f_{ \mathbf{k}} \rangle, \label{Dpp}
\end{eqnarray}
with all ${ \mathbf{q}}$ (${ \mathbf{k}}$) staying below (above) the Fermi surface. In Figs.\ref{fig_PTrTe_crossover}B-D, we plot $D_h$ and $D_p$ together in momentum space (as varying ${ \mathbf{q}}$ and ${ \mathbf{k}}$), while keeping ${\mathbf{q}_0}$ and ${ \mathbf{k}_0}$ fixed nearby the Fermi surface. In the polaron regime (Fig.\ref{fig_PTrTe_crossover}B), we can see that both $D_h$ and $D_p$ are extensively distributed in a broad angular range near the Fermi surface. In comparison, the correlation develops a visible crystalline feature in the trimer and tetramer regimes. Specifically, the crystallization displays as a diagonal structure in trimer regime (Fig.\ref{fig_PTrTe_crossover}C) and a regular triangle in tetramer regime (Fig.\ref{fig_PTrTe_crossover}D).  To trace the evolution, we further examine the particle-particle angular correlation, i.e., $\sum_{{ \mathbf{k}}, { \mathbf{k}}'}D_p({ \mathbf{k}},{ \mathbf{k}}')\delta_{\theta_{ \mathbf{k}}-\theta_{ \mathbf{k}'}, \theta_{{ \mathbf{k}}{ \mathbf{k}}'}}$ as a function of relative angle $\theta_{{ \mathbf{k}}{ \mathbf{k}}'}$, see Figs.\ref{fig_PTrTe_crossover}E-F. We can see that during the polaron-trimer crossover (Fig.\ref{fig_PTrTe_crossover}E), the angular dependence of $D_p$ becomes gradually pronounced at $\theta_{{\mathbf{k}}{ \mathbf{k}}'}=\pi$, signifying a diagonal correlation (similar to Figs.\ref{fig_PTr_crossover}C-D); while during the trimer-tetramer crossover (F), the peaks of $D_p$ gradually deviate from $\pi$ and finally end up at $2\pi/3$ and $4\pi/3$, signifying the emergence of triangular correlation. All these processes can be well captured by the general ansatz $P_7(0)$, which produces a considerably lower variational energy than $Tr_5(0)$ and $Te_4(0)$, see Fig.\ref{fig_PTr_crossover}E. Again, the energy crossing between $Tr_5(0)$ and $Te_4(0)$ does not stand for any real transition.

\begin{figure}[t]
\includegraphics[width=8.2cm]{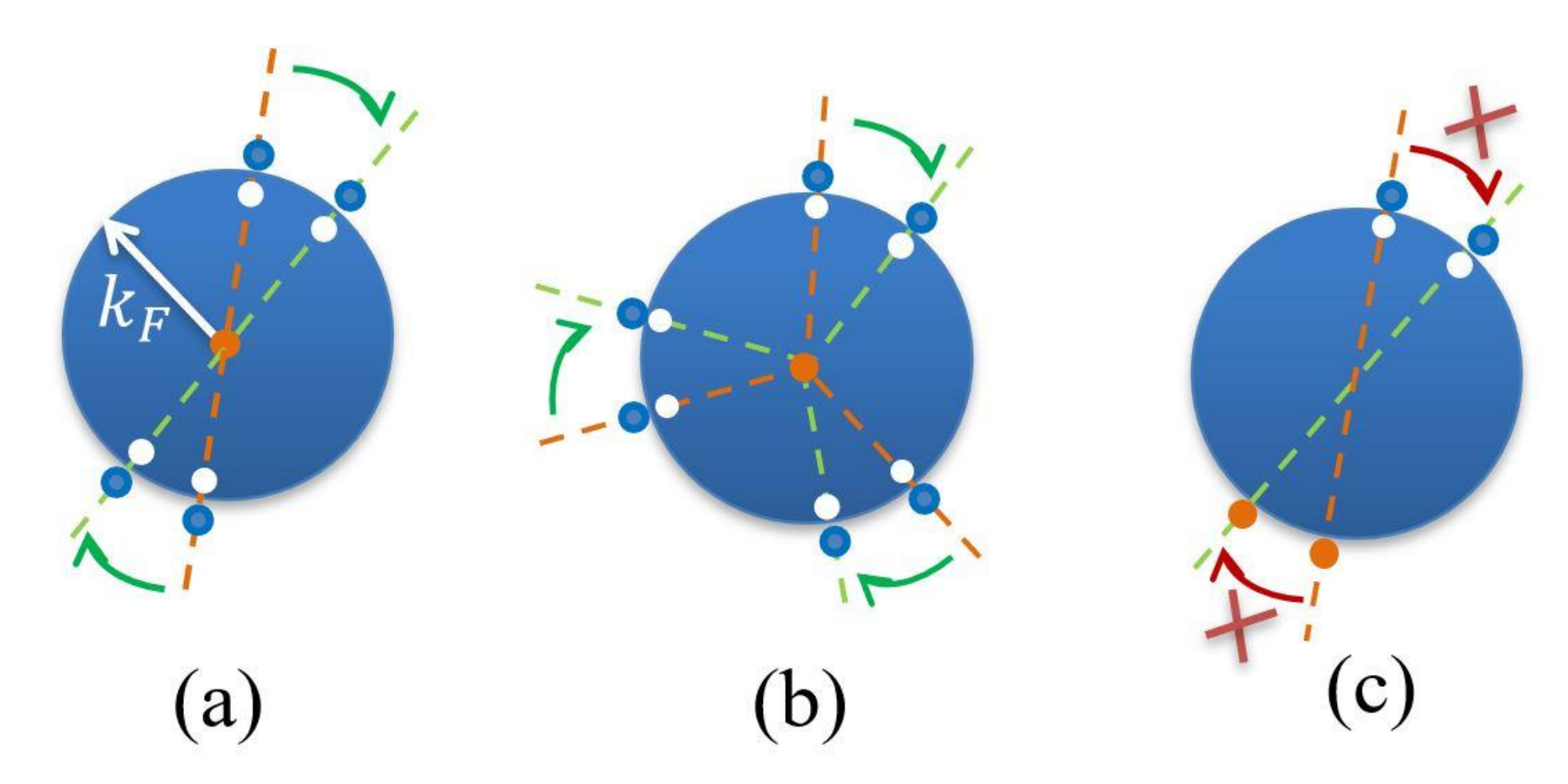}
\caption{(Color Online). Schematics of the crystalline particle-hole configurations and their mutual scattering. The blue and white points denote, respectively, the particle and hole of majority fermions, and the orange point denotes the impurity. In the trimer or tetramer regime, the scattering between different diagonal configurations (a) or between different triangular ones (b) is allowed by multiple scattering processes. All configurations in (a,b) are with the same total momentum ${\cp Q}=0$ (taking the Fermi sea $|FS\rangle_{\rm N}$ as reference). This is in contrast to the molecule case with $|{\cp Q}|=k_F$(c), where different p-h configurations cannot be scattered freely due to the conservation of total ${\cp Q}$.
} \label{fig_ph_scattering}
\end{figure}

In above, we have shown that the emergent crystalline correlations in trimer and tetramer regimes all display high symmetries, i.e., either with a diagonal or a regular triangular structure.  Physically, it is because these highly symmetric structures are associated with a large degeneracy manifold, in which they can be scattered freely to other similar configurations  to minimize the energy. As shown schematically in Figs.\ref{fig_ph_scattering}A-B, from a given p-h configuration with a diagonal or triangular structure near the Fermi surface, it can be coupled to another p-h configuration with the same structure via interactions, without changing the total momentum or costing any energy. This is quite similar to the facilitated three-body scattering between an impurity  and two fermions with Rashba spin-orbit coupling\cite{CuiYi}. With such high symmetry, the scattering phase space is maximized and the variational energy is expected to be significantly reduced. This is why the full $P_7(0)$ ansatz can produce reasonably lower energy than the trimer or tetramer ansatz, as shown in Fig.\ref{fig_PTr_crossover}E and 4G, since the latter just corresponds to a specific hole configuration in Figs.\ref{fig_ph_scattering}A-B with much smaller variational space. Moreover, we emphasize that the final state ($P_7(0)$) is a superposition of all these degenerate configurations, and thus the rotational invariance of the whole system can be recovered. This is how
the crystalline correlation can only be reflected in the two- and higher-body correlation functions, but not in the one-body density profile which obeys the rotational symmetry. In this sense, the emergent crystallization in high-order correlations can be regarded as a spontaneous symmetry breaking phenomenon.

In addition, it is important to note that the diagonal and regular triangular p-h distributions guarantee the system remain in ${ \mathbf{Q}}=0$ sector (take $|\mathrm{FS}\rangle_N$ as reference). This is quite different from the molecule state which belongs to the $|{ \mathbf{Q}}|=k_F$ sector (note that the ${ \mathbf{Q}}=0$ molecule  belongs to an excited manifold and see note S3 for more details). In this case, different hole configurations cannot be scattered freely due to the conservation of total ${ \mathbf{Q}}$, see also Fig.\ref{fig_ph_scattering}C. Such forbidden scattering is a crucial reason why the molecule ansatz (corresponding to a specific hole configuration in Fig.\ref{fig_ph_scattering}C) can energetically  well approximate the full $|{ \mathbf{Q}}|=k_F$ state in strong coupling regime, see Fig.\ref{fig_PMtransition}B.
To conclude, when the two-body (dimer) correlation dominates, the system undergoes a first-order transition from polaron (${ \mathbf{Q}}=0$) to molecule ($|{ \mathbf{Q}}|=k_F$); while when the three-body (trimer), four-body (tetramer) and even higher-body correlations dominate, the system just evolves smoothly and all these highly correlated states can be adiabatically approached from the polaronic regime within the same
${ \mathbf{Q}}=0$ sector (Fig.\ref{fig_PTr_crossover} and Fig.\ref{fig_PTrTe_crossover}).

Finally, we comment on the difference between the crystalline correlation in Fermi polarons and that in purely few-body systems.   In the latter case, the emergence of crystalline correlation only relies on the values of $\eta$ as compared to $\eta_{tr}$, $\eta_{te}$, etc,  with the crystalline radius changing solely with the scattering length\cite{Peng}.
In contrast, in Fermi polaron system, the crystalline correlation {\it gradually} emerges near the Fermi surface when the coupling  is strong enough. For instance, the  triangular  correlation appears only after a polaron-trimer-tetramer crossover (Fig.\ref{fig_PTrTe_crossover}), but not exists for arbitrary couplings as in the four-body system. In this sense, the majority Fermi sea serves as a many-body reservoir to output few-body correlations required for the Fermi polaron.  Moreover, the crystallization in Fermi polarons first appears near the Fermi surface, where the particle-hole excitations play an important role in its emergence. More details can be found in note S4 and Fig.S4.

\section{DISCUSSION}

In summary, we have demonstrated the dominant three- and four-body correlations in the mass-imbalanced Fermi polarons in 2D. As increasing the mass ratio or attraction strength between fermions and the impurity, the system is found to evolve smoothly from a polaronic to dressed trimer and tetramer states. Such a crossover is accompanied by the emergence of momentum-space crystallization of majority fermions,  as featured by the diagonal or regular triangular structure of p-h excitations nearby the Fermi surface.
The emergent crystallization can be detected through the density-density correlation function of majority fermions in momentum space, as having been successfully implemented in cold-atom systems using the atom noise in absorption images\cite{theory_noise,expt_noise1,expt_noise2,expt_noise3,expt_noise4,expt_noise5} or the single atom resolved images\cite{Jochim_expt1,Jochim_expt2}.
Since all the Fermi-Fermi mixtures realized so far\cite{K_Li1, K_Li2, K_Li3, Dy_K1,Dy_K2, Yb_Li, Yb_Li2,Cr_Li,Cr_Li2} have the mass ratio $\eta>\eta_{te}=3.38$\cite{Peng}, we expect that the polaron-trimer-tetramer crossover and its associated crystalline correlations can be readily observed in these Fermi polaron systems once confined in 2D. To achieve the effective 2D geometry,  a strong axial confinement can be applied in realistic quasi-2D setup with confinement frequency much larger than all relevant scales of the system ($E_F$, $E_{2b}$, etc), under which the effective interaction is solely described by the reduced 2D scattering length\cite{Petrov_2D}.

Given the robust intrinsic relation between polaron, trimer and tetramer, the conclusion in this work  can be generalized to Fermi polaron systems with arbitrarily high-order correlations and in higher dimension (3D). For instance, a ($1+4$) pentamer can emerge in 2D at a larger $\eta$\cite{Peng}, which is expected to drive a smooth crossover of Fermi polarons further from the dressed tetramer to pentamer regime as increasing the coupling strength.
Similar physics is also expected for 3D, namely, a polaron-molecule transition occurs when $\eta<\eta_{tr}=8.2$\cite{KM}, while a smooth polaron-trimer crossover with diagonal crystallization occurs when $\eta_{tr}<\eta<\eta_{te}$ and  a promising system for its observation is $^{53}$Cs-$^{6}$Li mixture\cite{Cr_Li,Cr_Li2}. Surely, the emergence of crystalline tetramer and pentamer correlations would be even more intriguing on a 3D Fermi surface, which would require even  larger $\eta$ as inferred by the cluster formation in few-body sector\cite{universalTetramer,pentemer}.  In comparison, the situation in 1D is quite different, where the polaron-molecule transition is absent even for the equal mass case\cite{McGuire, Guan, Cui2021}. In fact, in 1D the trimer emerges exactly at equal mass ($\eta=1$)\cite{1D_trimer}, and  one thus expects  no sharp transition for any $\eta>1$ and the ground state is always at $Q=0$. The higher-body correlations can also play an important role in 1D Fermi polarons in view of various cluster bound states supported at larger $\eta$\cite{Petrov_1D}.

Our results shed light on the quantum phases of highly polarized Fermi-Fermi mixtures with mass imbalance. First, the results suggest that if the majority fermions have a much heavier mass than the minorities, the dominant correlation can switch from two-body to three- and even higher-body sectors. As a result, the ground state is no longer a pairing superfluid as extensively studied in literature, but a new quantum state dominated by $n$-body($n\ge3$) correlation.
Secondly, the results suggest the absence of sharp phase transition or phase separation in the highly polarized Fermi-Fermi mixtures with large mass imbalance, where the evolution from the normal to the trimer- or tetramer-correlated   phases is a smooth crossover.
This is in contrast to the equal mass case which supports a first-order transition from the normal to pairing superfluid as well as a phase separation between them\cite{PS_theory}. In future, a precise description of such highly polarized $n$-body correlated state requires the knowledge of effective interactions within trimers, tetramers, and between trimer/tetramer and majority fermions.
Finally, the regime with intermediate polarization may exhibit even richer physics due to various competing orders associated with different few-body correlations. Our study thus opens an avenue for searching novel quantum phases beyond the traditional pairing superfluids in the mass-imbalanced fermion systems.

\section{EXPERIMENTAL PROCEDURES}
\subsection{Resource Availability}
\subsubsection{Lead contact}
Further information and requests for resources should be directed to and will be fulfilled by the lead contact, Xiaoling Cui (xlcui@iphy.ac.cn).

\subsubsection{Materials availability}
This study did not generate any new, unique reagents.

\subsubsection{Data and Code Availability}
All data and code are available from
the lead contact upon request.

\section{ACKNOWLEDGEMENTS}
We thank Meera Parish and Jesper Levinsen for helpful discussions and for pointing out the additional solution of polaron ansatz. % Beijng PARATERA Tech CO., Ltd. for providing HPC resources.
The work is supported by the National Key Research and Development Program of China (2018YFA0307600), the National Natural Science Foundation of China (12074419 and 12134015), and the Strategic Priority Research Program of Chinese Academy of Sciences (XDB33000000).

\section{AUTHOR CONTRIBUTIONS}
X.C. conceived the idea and supervised the project. R.L. performed the numerical calculations and C.P. provided assistance in numerics. All authors discussed the results and contributed to writing the paper.

\section{DECLARATION OF INTERESTS}
The authors declare no competing interests.

\clearpage
\begin{widetext}

\section*{SUPPLEMENTAL INFORMATION}
\setcounter{page}{1}
\setcounter{figure}{0}
\setcounter{equation}{0}
\renewcommand{\thefigure}{S\arabic{figure}}
\renewcommand{\theequation}{S\arabic{equation}}
\pagestyle{empty}

\subsection*{Note S1: Coupled integral equations for \bm{$P_7$}}

Given the $P_7({ \mathbf{Q}})$ ansatz Eq. 2 in the maintext, we rescale the coefficients as $\alpha^{(l)}\equiv \psi^{(l)}/\psi^{(0)}$ and define three %After simplification, the final coupled equations are
auxiliary functions fascilitated by the contact interaction:
\begin{eqnarray}
&&f(\mathbf{q})\equiv \frac{g}{S}(1+\sum_{\mathbf{k}} \alpha^{(1)}_{\mathbf{kq}}), \nonumber\\
&&G(\mathbf{k,q',q})\equiv \frac{g}{S}\sum_{\mathbf{k}'} (\alpha^{(2)}_{\mathbf{k'kq'q}}+\alpha^{(1)}_{\mathbf{kq'}}-\alpha^{(1)}_{\mathbf{kq}}),  \nonumber\\
&&h(\mathbf{k',k,q'',q',q})\equiv \frac{g}{S}\sum_{\mathbf{k}''} \alpha^{(3)}_{\mathbf{k''k'kq''q'q}}.
\end{eqnarray}
In this way, the original equations for $\psi^{(l)}$ ($l=0,1,2,3$)  obtained from the Schr{\"o}dinger equation can be reduced to the following four equations:
\begin{eqnarray}
\label{eq:iter1}
&&E-\epsilon_{\mathbf{Q}}^a=\sum_{\mathbf{q}} f(\mathbf{q}),\\
\label{eq:iter2}
&&[\frac{1}{g}-\sum_{\mathbf{k}_1}\frac{
1}{E_{\mathbf{k}_1\mathbf{q}_1}}]f(\mathbf{q}_1)=\frac{\sum_{\mathbf{q}} f(\mathbf{q})}{E-\epsilon_{\mathbf{Q}}^a}
-\sum_{\mathbf{k}_1\mathbf{q}}\frac{ G(\mathbf{k}_{1},\mathbf{q},\mathbf{q}_{1})}{E_{\mathbf{k}_1\mathbf{q}_1}},\\
\label{eq:iter3}
&&[\frac{1}{g}-\sum_{\mathbf{k}_{1}}\frac{1}
{E_{\mathbf{k}_{1}\mathbf{k}_{2}\mathbf{q}_{1}\mathbf{q}_{2}}}]G(\mathbf{k}_{2},\mathbf{q}_{1},\mathbf{q}_{2}) = \alpha^{(1)}_{\mathbf{k_2q_1}}- \alpha^{(1)}_{\mathbf{k_2q_2}} -\sum_{\mathbf{k}_{1}}\frac{1}
{E_{\mathbf{k}_{1}\mathbf{k}_{2}\mathbf{q}_{1}\mathbf{q}_{2}}}G(\mathbf{k}_{1},\mathbf{q}_{1},\mathbf{q}_{2})\nonumber\\
&&\quad\quad+\sum_{\mathbf{k}_{1}\mathbf{q}}\frac{h
(\mathbf{k}_{1},\mathbf{k}_{2},\mathbf{q}_{1},\mathbf{q}_{2},\mathbf{q})}{E_{\mathbf{k}_{1}\mathbf{k}_{2}\mathbf{q}_{1}\mathbf{q}_{2}}},\\
\label{eq:iter4}
&&[\frac{1}{g}-\sum_{\mathbf{k}_{1}}\frac{1}{E_{\mathbf{k}_{1}\mathbf{k}_{2}\mathbf{k}_{3}\mathbf{q}_{1}\mathbf{q}_{2}
\mathbf{q}_{3}}}]
h(\mathbf{k}_{2},\mathbf{k}_{3}, \mathbf{q}_{1}, \mathbf{q}_{2},\mathbf{q}_{3}) = \alpha^{(2)}_{\mathbf{k}_{2}\mathbf{k}_{3}\mathbf{q}_{1}\mathbf{q}_{2}}-\alpha^{(2)}_{\mathbf{k}_{2}\mathbf{k}_{3}\mathbf{q}_{1}
\mathbf{q}_{3}}+ \alpha^{(2)}_{\mathbf{k}_{2}\mathbf{k}_{3}\mathbf{q}_{2}\mathbf{q}_{3}}\nonumber\\
&&\quad\quad+\sum_{\mathbf{k}_{1}}\frac{1}{E_{\mathbf{k}_{1}\mathbf{k}_{2}\mathbf{k}_{3}\mathbf{q}_{1}
\mathbf{q}_{2}\mathbf{q}_{3}}}[h(\mathbf{k}_{1},\mathbf{k}_{2},\mathbf{q}_{1},\mathbf{q}_{2},\mathbf{q}_{3})
-h(\mathbf{k}_{1},\mathbf{k}_{3},\mathbf{q}_{1},\mathbf{q}_{2},\mathbf{q}_{3})],
\end{eqnarray}
Here the energies are defined as $E_{\mathbf{k_1...k_iq_1...q_i}}=E-\epsilon^a_{ \mathbf{P}}+\sum_{j=1}^i(\epsilon^f_{\mathbf{q_j}}-\epsilon^f_{\mathbf{k_j}})$, with ${ \mathbf{P}}={ \mathbf{Q}}+\sum_{j=1}^i(\mathbf{q_j}-\mathbf{k_j})$. The scaled coefficients can be represented by the auxiliary functions via
\begin{eqnarray}
\alpha^{(1)}_{\mathbf{k}_1\mathbf{q}_1}&=&\frac{
f(\mathbf{q}_1)-\sum_{\mathbf{q}} G(\mathbf{k}_{1},\mathbf{q},\mathbf{q}_{1})}{E_{\mathbf{k}_1\mathbf{q}_1}},\\
\alpha^{(2)}_{\mathbf{k}_{1} \mathbf{k}_{2} \mathbf{q}_{1} \mathbf{q}_{2}} &=&\frac{1}{E_{\mathbf{k}_{1}\mathbf{k}_{2}\mathbf{q}_{1}\mathbf{q}_{2}}}[G(\mathbf{k}_{2},\mathbf{q}_{1},\mathbf{q}_{2})
-G(\mathbf{k}_{1},\mathbf{q}_{1},\mathbf{q}_{2})+\sum_{\mathbf{q}}h(\mathbf{k}_{1},\mathbf{k}_{2},
\mathbf{q}_{1},\mathbf{q}_{2},\mathbf{q})],\\
\alpha^{(3)}_{\mathbf{k}_{1} \mathbf{k}_{2}\mathbf{k}_{3} \mathbf{q}_{1} \mathbf{q}_{2}\mathbf{q}_{3}} &=&\frac{1}{E_{\mathbf{k}_{1}\mathbf{k}_{2}\mathbf{k}_{3}\mathbf{q}_{1}\mathbf{q}_{2}\mathbf{q}_{3}}}
[h(\mathbf{k}_{1},\mathbf{k}_{2},\mathbf{q}_{1},\mathbf{q}_{2},\mathbf{q}_{3})
-h(\mathbf{k}_{1},\mathbf{k}_{3},\mathbf{q}_{1},\mathbf{q}_{2},\mathbf{q}_{3})\nonumber\\
&&+h(\mathbf{k}_{2},\mathbf{k}_{3},\mathbf{q}_{1},\mathbf{q}_{2},\mathbf{q}_{3})].
\end{eqnarray}

It is noted that above equations can well reproduce the trimer and tetramer bound states in vacuum by sending $k_F$ to zero. For instance, by only keeping $G({ \mathbf{k}},{ \mathbf{q}}_1,{\mathbf{q}}_2)\rightarrow G({ \mathbf{k}})$ terms, Eq. \ref{eq:iter3} can be reduced to the equation of trimer bound state. Similarly, the tetramer binding energy can be extracted from Eq. \ref{eq:iter4} by only keeping $h(\mathbf{k}_{1},\mathbf{k}_{2},\mathbf{q}_{1},\mathbf{q}_{2},\mathbf{q}_{3})\rightarrow h(\mathbf{k}_{1},\mathbf{k}_{2})$ terms.

\subsection*{Note S2: Numerical details for \bm{$P_7$} ansatz}

In principle, the coupled Eqs. \ref{eq:iter1}-\ref{eq:iter4} can be transformed as matrix equation by discretizing each momentum $\mathbf{q}$ and $\mathbf{k}$. The energy and wavefunction can then be obtained by directly solving the matrix equation. However, the numerical cost can be very large. We first analyze the complexity of directly solving a large matrix equation for $P_7$.
If we perform the discretization $7-13-8-13$ with respect to $q-\theta_q-k-\theta_k$ in the polar coordinates, the dimension of the corresponding matrix will be $\sim10^{10}\times10^{10}$. Though the dimension can be reduced by utilizing the antisymmetry (due to Fermi statistics) and rotational symmetry of the wavefunctions, the direct solution of the matrix equation is still out of reach.

Instead of solving a large matrix equation, we obtain the energy and wavefunctions by iteratively solving the coupled Eqs.\ref{eq:iter1}-\ref{eq:iter4}. Specifically, we first set the initial wavefunctions $f_i$, $g_i$ $h_i$ and energy $E_i$. Then by substituting them into the right-hand side of the coupled Eqs. \ref{eq:iter1}-\ref{eq:iter4}, we can obtain new wavefunctions $f_f$, $g_f$ $h_f$ and energy $E_f$, which will be treated as the updated initial inputs. In addition, to avoid divergency during the iteration, we introduce a step factor $s$ to control the changes between successive iterations, i.e., we take $(1-s)f_i+sf_f\rightarrow f_i$, $(1-s)g_i+sg_f\rightarrow g_i$, $(1-s)h_i+sh_f\rightarrow h_i$ and $(1-s)E_i+sE_f\rightarrow E_i$. The iterations stop once the changes of wavefunctions and energies between successive iterations are smaller than the criterions we set. In Fig.S1, we show an example for the numerical process of the iteration.

In our numerics, we have also ensured the convergence of the results with respect to different momentum cutoffs ($k_c$) and discretization scheme.  In Fig.S2, we show the energy and weight distributions for mass ratio $\eta=3$ by choosing different $k_c$ and different momentum discretizations, which show agreement to a good extent.

\subsection*{Note S3: An additional solution from \bm{$P_7(0)$}}

Interestingly, in the strong coupling regime $P_7(0)$ can produce another type of solution that has very close energy with $M_4(0)$. This solution is very different from the one that is adiabatically evolved from the polaronic state in the weak coupling regime, and thus cannot be approached adiabatically from the weak coupling regime. As shown in the inset of Fig.S3(A), this state involves a special type of p-h excitations on the Fermi surface, namely, two holes (${ \mathbf{q}}_1,\ {\mathbf{q}}_2$) distribute evenly in each side of the particle (${\mathbf{k}}$) with relative angle $\theta_{{ \mathbf{q}}_1{ \mathbf{k}}}=\theta_{{ \mathbf{q}}_2{ \mathbf{k}}}=\pi/3$, which gives ${ \mathbf{q}}_1+{ \mathbf{q}}_2-{ \mathbf{k}}=0$. In this way, if we take the $(N-1)$ Fermi sea state in the molecule ansatz as $|\mathrm{FS}\rangle_{N-1}=f^{\dag}_{ \mathbf{k}} f_{\mathbf{q}_1}f_{ \mathbf{q}_2}|\mathrm{FS}\rangle_N$,  the molecule ansatz $M_{2n+2}(0)$ can be well covered by $P_{2n+5}(0)$. Namely, the latter  can reproduce the former by setting $\psi^{(0)}=\psi^{(1)}=0$ and requiring the rest high p-h excitations share the same structure as illustrated above.

In our numerics, we have obtained such a solution from $P_7(0)$ ansatz by choosing a special initial state at the beginning of iterative evolution, which incorporates both the molecule feature (we have taken $M_2(0)$ for simplicity) and the special structure of additional p-h excitations. To distinguish different solutions, from now on we call this state as $P_7(0,M)$, and the one adiabatically evolved from the weak-coupling polaronic state as $P_7(0,P)$. In Fig.S3(A-C), we show the energies of different ansatz and the properties of $P_7(0,M)$ near the polaron-molecule transition for the case of $\eta=2$. Three remarks on $P_7(0,M)$ are as follows:

(i) $P_7(0,M)$ is orthogonal to $P_7(0,P)$ and they have a level crossing at certain point. As shown in Fig.S3(A), the energy of $P_7(0,M)$ closely follow $M_4(0)$ and crosses the energy of $P_7(0,P)$ near the polaron-molecule transition point. The negligible overlap between $P_7(0,M)$ and $P_7(0,P)$, as shown in Fig.S3(B), signifies the orthogonality of the two states and confirms the level crossing behavior.

(ii) $P_7(0,M)$ can only be supported in the strong coupling regime beyond certain interaction strength. For the case of $\eta=2$, a  convergent solution of $P_7(0,M)$ exists for $\ln(k_Fa_{2d})\leq-1.1$, but not for weaker couplings (Fig.S3(C)).  For instance, we have checked that at $\ln(k_Fa_{2d})=-0.9$, no convergent energy can be found to be close to $M_4(0)$, and the iteration can evolve to states that are very different from the initial one and the energies quickly depart from that of $M_4(0)$. In this case, $P_7(0,M)$ is no longer locally stable in the large scattering  phase space, and the iteration will finally result in the true ground state $P_7(0,P)$.

(iii) In the regime where $P_7(0,M)$ is stabilized, it always has a higher energy than $P_7(k_F)$ (see Fig.S3(A)). Physically, this is because in the strong coupling limit $P_7(k_F)$ approaches $M_6(0)$ while $P_7(0,M)$ approaches $M_4(0)$; clearly, $M_4(0)$ is always energetically unfavorable than $M_6(0)$ due to the lower level of p-h excitations. Therefore, the ground state in strong coupling limit, up to three p-h excitations, is given by $P_7(k_F)$ instead of $P_7(0,M)$. In this way, the polaron-molecule transition is still present and still given by the energy crossing between $P_7(0)$ and $P_7(k_F)$.

To summarize, (i) and (ii) shows that a stable $P_7(0,M)$ cannot be approached adiabatically from the polaronic state in weak coupling limit, and it can only survive in the strong coupling regime by molecule formation. (iii) shows that with a fixed truncated level of p-h excitations on top of $|\mathrm{FS}\rangle_N$, the lowest molecule state is given by $P_{2n+1}(k_F)$, rather by a special solution of $P_{2n+1}(0)$ (in the current case $n=3$).   Because of all above, the presence of $P_7(0,M)$  does not affect the occurrence and the nature of polaron-molecule transition in Fermi polaron system with small mass ratios.

\subsection*{Note S4: Comparison of crystalline correlation between Fermi polarons and few-body systems}

The crystalline correlation in Fermi polarons is very different from that in pure few-body systems. As a concrete example, in Fig.S4 we take the mass ratio $\eta=40/6$ and show the difference of tetramer correlations in Fermi polaron and in $1+3$ system.

As show in Fig.S4(A-D) for the few-body system, the crystalline triangular correlation of heavy fermions exists for all couplings, with the crystalline radius changing solely with $a_{2d}$.
%, see Fig.\ref{fig_few_body} (a1-a4) for $1+3$ system with $\eta=40/6$. %Fig.\ref{fig_few_body}(a) for the density correlation of four-body system with $\eta=40/6$. %, the crystallized tetramer correlation is there for all coupling strengths, with the crystalline radius changed with couplings.
In comparison, the triangular correlation in Fermi polaron {\it gradually} emerges after the polaron-trimer-tetramer crossover as increasing the coupling strength, see Fig.S4(E-H). Moreover, the triangular structure first shows up near the Fermi surface, indicating an important role of the Fermi sea background and the particle-hole excitations therein for its emergence. All these features are distinct from the few-body case. However, for extremely strong coupling, such as in Fig.S4(D,H), the correlation profiles are essentially  identical between the few-body and polaron systems, due to the very deep tetramer bound state and a much larger crystalline radius  $(\gg k_F)$.  In this case the Fermi sea plays little roles in affecting the correlation profiles.

\clearpage

\begin{figure}[h]
\centering
\includegraphics[width=0.6\textwidth]{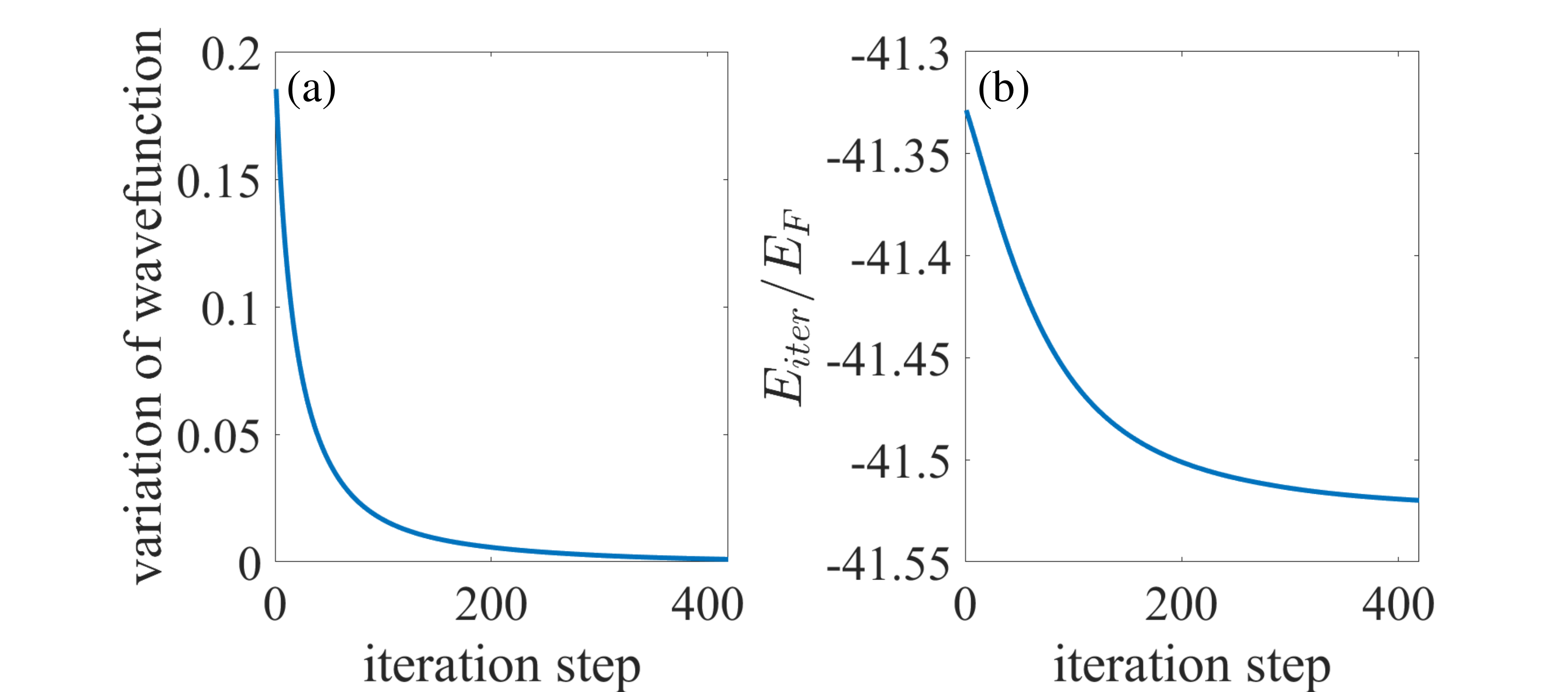}
\caption{ \label{fig:iteration}\textbf{Details of the iterations for mass ratio \bm{$\eta=3$} and interaction strength \bm{$\textrm{ln}(k_Fa_{2d})=-1.15$}.} (\textbf{A}) Changes of wavefunctions between successive iterations. (\textbf{B}) Energy evolution during the iteration. The discretization is taken as $7-13-8-13$ with respect to $q-\theta_q-k-\theta_k$ in the polar coordinates and the cutoff of momentum is taken as $k_c=50k_F$.}
\end{figure}

\begin{figure}[h]
\centering
\includegraphics[width=0.7\textwidth]{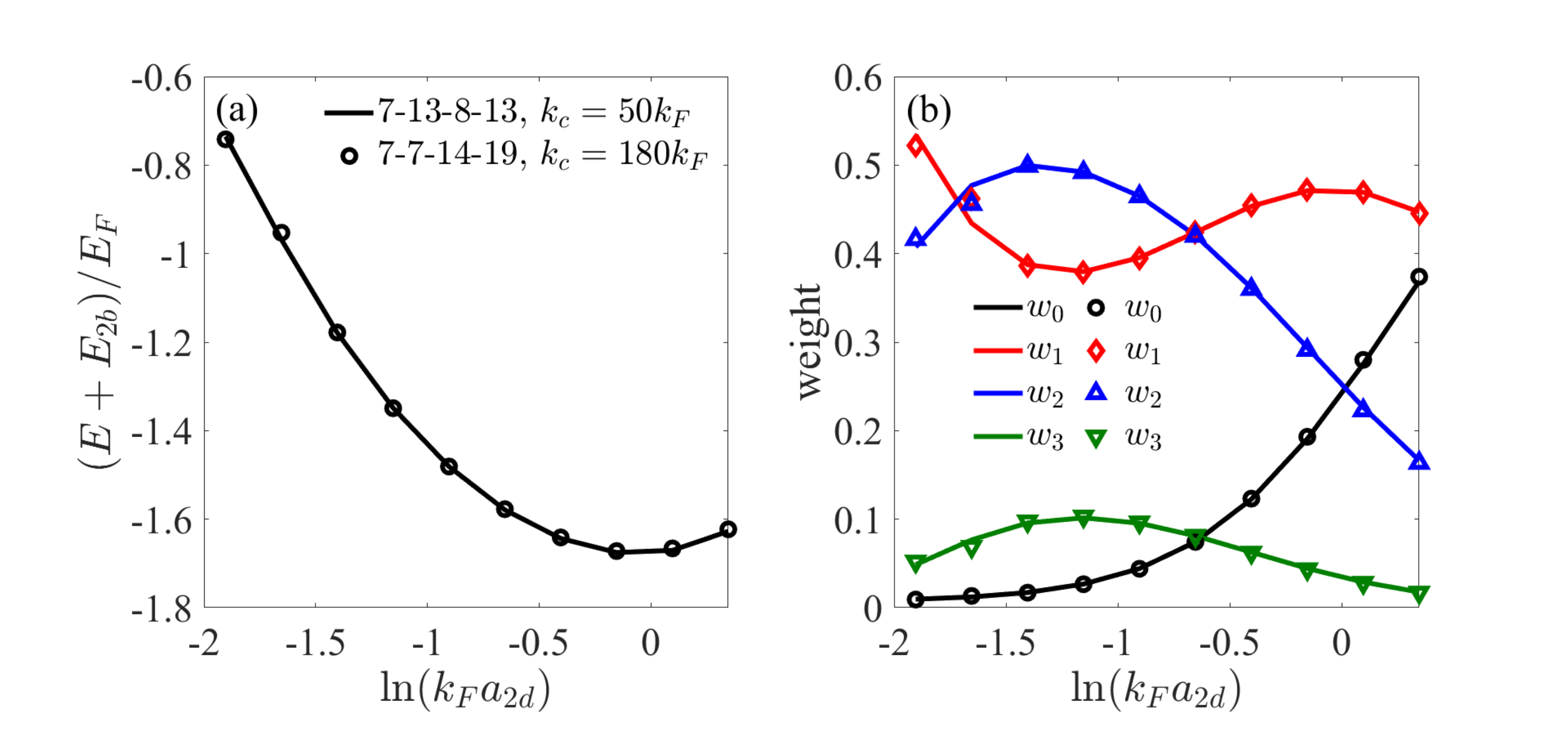}
\caption{ \label{fig:Eweight}\textbf{Convergency of the results for mass ratio \bm{$\eta=3$}.} (\textbf{A}) Energies obtained from discretization $7-13-8-13$ and $7-7-14-19$ with respect to $q-\theta_q-k-\theta_k$ in the polar coordinates. The cutoff of momentum is taken as $k_c=50k_F$ and $k_c=180k_F$, respectively. (\textbf{B}) Weight distributions with discretization and momentum cutoff the same as in (\textbf{A}). Lines: $7-13-8-13$ and $k_c=50k_F$; Symbols: $7-7-14-19$ and $k_c=180k_F$. }
\end{figure}

\begin{figure}[h]
\includegraphics[width=1\textwidth]{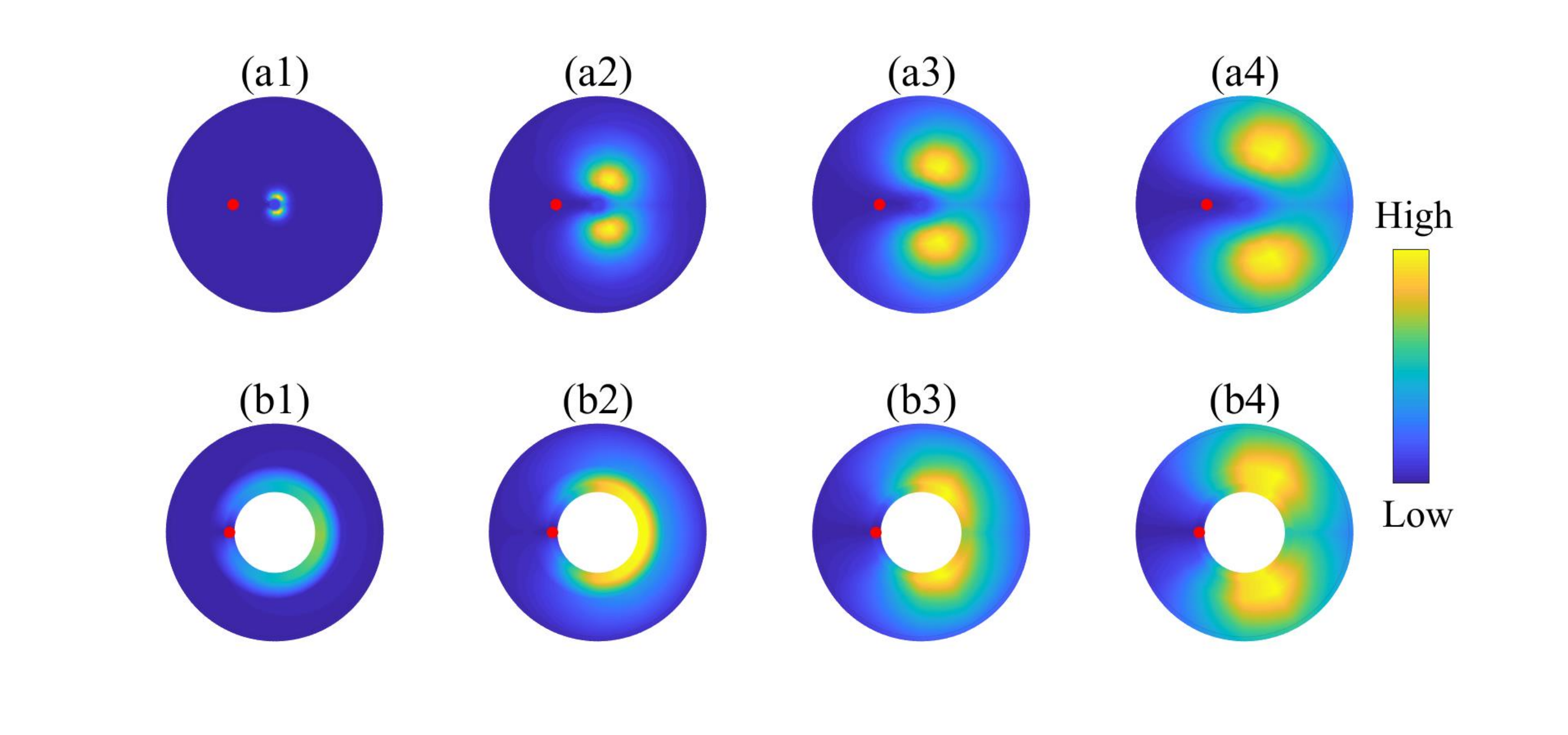}
\caption{ \label{fig:excited} \textbf{Additional solutions from \bm{$P_7(0)$}.} (\textbf{A}) Energy comparison between $P_7(0,M)$, $P_7(0,P)$, $P_7(k_F)$ and $M_4(0)$ in the regime near polaron-molecule transition for $\eta=2$. Note that $P_7(0,P)$ is the same as $P_7(0)$ in the main text, which adiabatically evolves from the polaronic state in weak coupling limit. To obtain $P_7(0,M)$, we have chosen a very different initial state to start the numerical iteration(see text). (\textbf{B}) Overlap between $P_7(0,M)$ and $P_7(0,P)$. In the stabilized regime of $P_7(0,M)$, its overlap with $P_7(0,P)$ is below $10^{-13}$ and negligible. (\textbf{C}) Energy evolution in the iteration process of $P_7(0,M)$ for two typical coupling strengths. The convergent result is obtained at $\textrm{ln}(k_Fa_{2d})=-1.1$ but not at $\textrm{ln}(k_Fa_{2d})=-0.9$. The energies of $M_4(0)$ for both cases are marked by horizontal gray lines.}
\end{figure}

\begin{figure}[h]
\centering
\includegraphics[width=15cm]{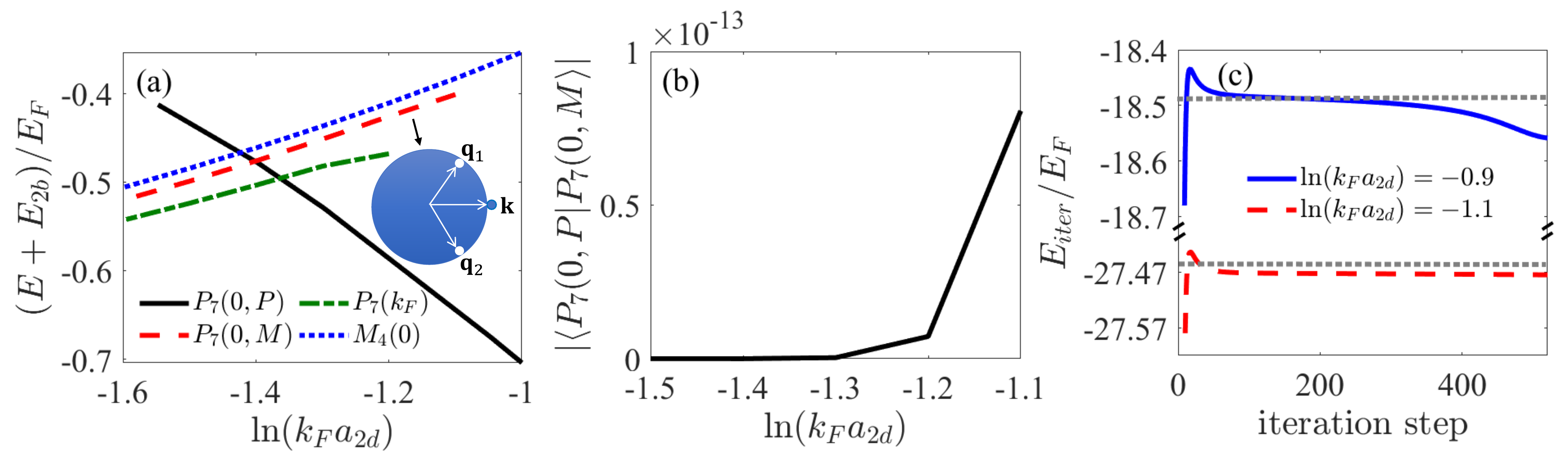}
\caption{\textbf{Comparison of particle-particle correlation,} \bm{$D_p({ \mathbf{k}_0},{ \mathbf{k}})$}, \textbf{between the pure four-body system (\textbf{A}-\textbf{D}) and the Fermi polaron system (\textbf{E}-\textbf{H}) at mass ratio \bm{$\eta=40/6$}.} The coupling strengths are $\ln(k_Fa_{2d})=1.272$(\textbf{A},\textbf{E}), $0.172$(\textbf{B},\textbf{F}), $-0.328$(\textbf{C},\textbf{G}), $-0.728$(\textbf{D},\textbf{H}). Here we take a fixed ${ \mathbf{k}}_0=-1.08 k_F{ \mathbf{e}}_x$ (as marked by red circle) and show the contour plot of $D_p$ in ${ \mathbf{k}}$ space. In (\textbf{A}-\textbf{D}), $k_F$ is just a constant momentum unit without any physical meaning; the triangular correlation appears for all couplings and its radius is solely determined by the scattering length. In (\textbf{E}-\textbf{H}), the white round area denotes the Fermi sea; the triangular correlation is absent for weak couplings and shows up gradually from the polaron(\textbf{E})-trimer(\textbf{F})-tetramer(\textbf{G}) crossover. }
\end{figure}

\end{widetext}

\end{document}